\documentclass[aps,twocolumn]{revtex4}
\usepackage{amsmath,amssymb}
\usepackage{graphicx}% Include figure files
\usepackage{dcolumn}% Align table columns on decimal point
\usepackage{bm}% bold math
\begin{document}

\title{Will we observe black holes at LHC?}

\author{Marco Cavagli\`a}
\altaffiliation{Email: marco.cavaglia@port.ac.uk}
\affiliation{Institute of Cosmology and Gravitation, University of
Portsmouth, Portsmouth P01 2EG, U.K.}
\author{Saurya Das}
\altaffiliation{Email: saurya@math.unb.ca}
\affiliation{Department of Mathematics and Statistics, University
of New Brunswick, Fredericton, New Brunswick E3B 5A3, Canada}
\author{Roy Maartens}
\altaffiliation{Email: roy.maartens@port.ac.uk}
\affiliation{Institute of Cosmology and Gravitation, University of
Portsmouth, Portsmouth P01 2EG, U.K.~}

\date{\today}
\begin{abstract}

The generalized uncertainty principle, motivated by string theory
and non-commutative quantum mechanics, suggests significant
modifications to the Hawking temperature and evaporation process
of black holes. For extra-dimensional gravity with Planck scale
O(TeV), this leads to important changes in the formation and
detection of black holes at the the Large Hadron Collider. The
number of particles produced in Hawking evaporation decreases
substantially. The evaporation ends when the black hole mass is
Planck scale, leaving a remnant and a consequent missing energy of
order TeV.  Furthermore, the minimum energy for black hole
formation in collisions is increased, and could even be increased
to such an extent that no black holes are formed at LHC energies.

\end{abstract}


\maketitle

The possible existence of large extra dimensions
(LEDs)~\cite{Arkani-Hamed:1998rs} has opened up new and exciting
directions of research in quantum gravity. In this scenario,
standard-model particles are confined to the observable
4-dimensional ``brane" universe, whereas gravitons can access the
whole $d$-dimensional ``bulk" spacetime, being localized at the
brane at low energies. The effects of LEDs show up in new particle
states at energies below the fundamental gravitational scale 
\cite{giudice} and
in nonperturbative effects in the trans-Planckian energy regime
(for a review see~\cite{Cavaglia:2002si}.) The fundamental Planck
scale in LED models may be as small as a TeV, within reach of
near-future experiments. This leads to many interesting
possibilities in ``quantum gravity phenomenology''. One of the
most exciting quantum gravity signatures would be production and
decay of black holes (BHs) and other extended objects in cosmic
ray air-showers and at particle colliders such as the Large Hadron
Collider (LHC)~\cite{Giddings:2001bu,ac,dec}.

Hawking evaporation provides the observable signature of BH
formation at the TeV energy scale. After formation, BHs are
expected to lose the hair associated with multipole and angular
momenta, decay via Hawking radiation, and eventually either
disappear completely or leave a Planck-sized remnant. The final
state of BH decay is presently not understood, being in the realm
of quantum gravity theory. Here we investigate the implications
for BH decay of the generalized uncertainty principle (GUP).

It is commonly believed that quantum gravity implies the existence
of a minimum length~\cite{Garay:1994en}. This in turn leads to a
modification of the quantum mechanical uncertainty principle:
\begin{equation}
\Delta x_i \gtrsim \frac{\hbar}{\Delta p_i}\left[1+\left(\alpha'
\ell_{\text{Pl}} \frac{\Delta p_i}{\hbar}\right)^2\right]\,,
\label{gup1}
\end{equation}
where $\ell_{\text{Pl}}$ is the Planck length and $\alpha'$ is a
dimensionless constant of order one which depends on the details
of the quantum gravity theory. Equation~(\ref{gup1}) can be
derived in the context of non-commutative quantum
mechanics~\cite{Maggiore:1993zu}, string
theory~\cite{Amati:1988tn} or from minimum length
considerations~\cite{Maggiore:rv}. In the classical limit, $\Delta
x_i\gg \ell_{\text{Pl}}$, we recover the standard
Heisenberg uncertainty principle $\Delta x \Delta p\gtrsim
\hbar$. The correction term in Eq.~(\ref{gup1}) becomes effective
when momentum and length scales are near the Planck scale.
Equation~(\ref{gup1}) implies a minimum length scale:
\begin{equation}
\Delta x_i\gtrsim \Delta x_{\text{min}}\equiv 2\alpha'
\ell_{\text{Pl}}\,.
\end{equation}
The Hawking thermodynamical quantities can be derived
heuristically by applying the uncertainty principle to
4-dimensional BHs, and the GUP leads to modified thermodynamical
quantities~\cite{Adler:2001vs}. Here we generalize this to
$d$-dimensional BHs, which are relevant for BH production at TeV
scales.

A $d$-dimensional Schwarzschild BH has gravitational potential
$\Phi\propto(r_s/r)^{d-3}$, where the horizon radius is given by
\begin{equation}\label{rs}
r_s=\omega_d \ell_{\text{Pl}}m^{1/(d-3)}\,,~m={M\over
M_{\text{Pl}}}\,.
\end{equation}
Here $M$ is the mass and the dimensionless area factor is
$\omega_d = \{16\pi/[(d-2)\Omega_{d-2}]\}^{1/(d-3)}$, where
$\Omega_{d-2}$ is the area of $S^{d-2}$. Since Hawking radiation
is a quantum process, the ``emitted'' quanta must satisfy the
uncertainty principle. By modelling a black hole as a
$(d-1)$-dimensional cube of size $2r_s$, the uncertainty in the
coordinate of a massless Hawking particle at emission is estimated
as $\Delta x \sim 2r_s$. The true value is $\Delta x = 2Kr_s$,
where $K$ is a correction factor of order one that can in
principle be calculated for the spherical geometry of the horizon.
The Hawking temperature $T$ can be identified up to normalization
with the energy uncertainty of the emitted particles, $T\sim\Delta
E\sim c\Delta p$ (the Boltzmann constant is set to one). This
leads to the formula for the BH temperature~\cite{Adler:2001vs}.
If one uses the standard uncertainty relation, one deduces the
standard formula~\cite{Adler:2001vs}, and this is readily
generalized to recover the $d$-dimensional Hawking
temperature~\cite{Argyres:1998qn}
\begin{equation}
T_0=\left(\frac{d-3}{4\pi\omega_d}\right)M_{\text{Pl}}c^2 \,
m^{1/(3-d)}\,. \label{temp2}
\end{equation}
For the GUP case, $ \Delta p= (2\hbar/\Delta x) [1+
\sqrt{1-4\ell_{\text{Pl}}^2\alpha'^2/\Delta x^2}]^{-1}$ and
$\Delta x = 2Kr_s$. This leads to
\begin{equation}
T=2T_0\left(1+\sqrt{1-\frac{\alpha^2}{\omega_d^2
m^{2/(d-3)}}}\right)^{-1}\,, \label{temp1}
\end{equation}
where $\alpha=\alpha'/K$. Equation~(\ref{temp1}) generalizes the
4-dimensional result of Ref.~\cite{Adler:2001vs} to $d$
dimensions, and also to incorporate the GUP parameter $\alpha'$
and the geometrical correction factor $K$. In
Ref.~\cite{Adler:2001vs}, $\alpha=1$.

From Eq.~(\ref{temp1}) it is evident that GUP-quantum gravity
effects increase the characteristic temperature of the BH.
Therefore, we expect quantum BHs to be {\em hotter, shorter-lived
and with a smaller entropy} than semiclassical BHs of the same
mass. The BH temperature is undefined for $M<M_{\text{min}}$,
where
\begin{equation}
M_{\text{min}}= \frac{d-2}{8\Gamma\left(\frac{d-1}{2}\right)}
\left(\alpha\sqrt{\pi}\right)^{d-3} \, M_{\text{Pl}}\,.
\label{minmass}
\end{equation}
BHs with mass less than $M_{\text{min}}$ do not exist, since their
horizon radius would fall below the minimum allowed length. Hence
Hawking evaporation must stop once the BH mass reaches $M_{\rm
min}$. This can be shown by calculating the mass loss during
evaporation. The energy radiated per unit time is governed by the
Stefan-Boltzmann law. Following Ref.~\cite{Emparan:2000rs}, we
assume that the radiation takes place mainly along the
4-dimensional brane. (We neglect the energy losses due to emission
of gravitons into the bulk.) Since the surface gravity is constant
over the horizon, the Hawking temperature of the
higher-dimensional BH and that of the induced BH on the brane are
identical. Thus Eq.~(\ref{temp1}) can be used to calculate the
emission rate. The mass loss for a BH emitting on an
$n$-dimensional brane is given by~\cite{cavaglia}
\begin{equation}
\frac{dm}{dt}=-\frac{1}{cM_{\text{Pl}}}\bar\sigma_n A(n) T^n\,,
\label{massloss1}
\end{equation}
where $A(n)$ is the area of the induced BH on the brane and
$\bar\sigma_n$ is the effective $n$-dimensional Stefan-Boltzmann
constant,
\begin{equation}
\bar\sigma_n=\frac{\Omega_{n-3}\Gamma(n)\zeta(n)}{(2\pi\hbar
c)^{n-1}(n-2)} \sum_i c_i(n)\Gamma_{s_{i}}(n)f_i(n)\,. \label{sb}
\end{equation}
Here the sum is over all particle flavours, $c_i$ are the
$n$-dimensional degrees of freedom of the individual species, the
$\Gamma_{s_{i}}$'s are the $n$-dimensional greybody
factors~\cite{Kanti:2002nr}, and $f_i(n)=1$ or $1-2^{1-n}$ for
bosons or fermions. The area of the induced BH on the brane
is taken as the optical area in $n-2$ dimensions and is given
by
\begin{equation}
A(n)=\Omega_{n-2}r_c^{n-2}\,, \label{optarea}
\end{equation}
where $r_c=\left[(d-1)/2\right]^{1/(d-3)}
\left[(d-1)/(d-3)\right]^{1/2}r_s$ is the optical
radius~\cite{Emparan:2000rs}. Using Eq.~(\ref{temp1}) the BH rate
of mass loss on a 4-dimensional brane can be written as
\begin{equation}
\frac{dm}{dt}=16\left(\frac{dm}{dt}\right)_0
\left(1+\sqrt{1-\frac{\alpha^2}{\omega_d^2
m^{2/(d-3)}}}\right)^{-4}\,, \label{massloss2}
\end{equation}
where the standard rate is
\begin{equation}
\left(\frac{dm}{dt}\right)_0=-\frac{\mu}{t_{\text{Pl}}}
m^{-2/(d-3)}\,, \label{massloss3}
\end{equation}
with
\begin{equation}
\mu=\frac{\pi^2}{120}\frac{(d-1)^{(d-1)/(d-3)}}
{2^{2/(d-3)}\omega_d^2} \left(\frac{d-3}{4\pi}\right)^3\sum_i
c_i(4)\Gamma_{s_{i}}(4)f_i(4)\,. \label{mu}
\end{equation}
As expected, the GUP-corrected mass loss is larger than the
standard Hawking mass loss. Equation~(\ref{massloss2}) can be
integrated to give the decay time,
\begin{equation}
\tau=\left(\frac{d-3}{16\mu}\right)\left(\frac{\alpha}
{\omega_d}\right)^{d-1}I(4,d-6,\omega_d
m^{1/(d-3)}/\alpha)\,t_{\text{Pl}}\,, \label{tau1}
\end{equation}
where
\begin{equation}
I(m,n,x)=\int_1^x dz\, z^n\left(z+\sqrt{z^2-1}\right)^m\,.
\label{I}
\end{equation}
(This integral can be solved analytically.) Setting $\alpha=0$ we
obtain the standard Hawking decay time in $d$ dimensions,
\begin{equation}
\tau_0=\frac{1}{\mu}\left(\frac{d-3}{d-1}\right)
m^{(d-1)/(d-3)}\,t_{\text{Pl}} \,. \label{tau2}
\end{equation}
The BH entropy is
\begin{equation}
S=2\pi\omega_d\left(\frac{\alpha}{\omega_d}\right)^{d-2}
I(1,d-4,\omega_d m^{1/(d-3)}/\alpha)\,, \label{entropy1}
\end{equation}
and is always smaller than the standard ($\alpha=0$) Hawking
value,
\begin{equation}
S_0=\left(\frac{4\pi\omega_d}{d-2}\right)m^{(d-2)/(d-3)}\,.
\label{entropy2}
\end{equation}

From Eq.~(\ref{massloss3}) and Eq.~(\ref{tau2}), it is evident
that the final stage of standard Hawking evaporation is
catastrophic. The BH reaches in a finite time a stage with zero
mass, infinite radiation rate and infinite temperature. By
contrast, in the GUP scenario the existence of a minimum length
prevents the mass becoming smaller than $M_{\text{min}}$,
Eq.~(\ref{minmass}). At this point the emission rate,
Eq.~(\ref{massloss2}), becomes imaginary. Since the emission rate
is finite at the end, it could be argued that the BH decays
non-thermally by emitting a hard Planck-mass quantum in a finite
time  ${\cal O}(t_{\text{Pl}})$, once the final stage of
evaporation has been reached. However, the specific heat, ${\cal
C}=T\partial S/\partial T$, is
\begin{eqnarray}
{\cal C}&=&-2\pi\omega_d m^{(d-2)/(d-3)}\sqrt{1-
\frac{\alpha^2}{\omega_d^2 m^{2/(d-3)}}}\times\nonumber
\\ &&~~~~{}\times \left(1+ \sqrt{1-
\frac{\alpha^2}{\omega_d^2 m^{2/(d-3)}}}\right), \label{sh}
\end{eqnarray}
and vanishes at the endpoint, so that the BH cannot exchange heat
with the surrounding space. Thus the endpoint of Hawking
evaporation in the GUP scenario is characterized by a {\em
Planck-sized remnant} with maximum temperature,
\begin{equation}
T_{\text{max}}=2T_0\Big|_{M=M_{\text{min}}}\,.
\end{equation}

The way that the GUP prevents BHs from evaporating completely is
similar to the way that the standard uncertainty principle
prevents the hydrogen atom from collapsing. The existence of a
remnant as a consequence of the GUP was pointed in
Ref.~\cite{Adler:2001vs}, in the context of primordial BHs in
cosmology. Primordial BH remnants are possible candidates for dark
matter. Remnants have also been predicted in certain 
models of quantum black holes \cite{bdk}.

The multiplicity of a particle species $i$ produced in BH decay
is~\cite{cavaglia}
\begin{equation}
N_i=N\,\frac{c_i\Gamma_{s_i}f_i(3)}{\sum_j
c_j\Gamma_{s_{j}}f_j(3)}\,, \label{multii}
\end{equation}
where the total multiplicity $N$ is
\begin{equation}
N=\frac{30\zeta(3)}{\pi^4}\,S\,\frac{\sum_i c_i\Gamma_{s_i}f_i(3)}
{\sum_j c_j\Gamma_{s_{j}}f_j(4)}\,. \label{multin}
\end{equation}

The observable effects of the GUP at particle colliders are
related to the existence of a minimum mass and to the multiplicity
of the decay. We assume that the parton-parton center-of-mass (CM)
energy $E_{\text{cm}}$ is larger than $M_{\text{min}}$. Since the
GUP-corrected entropy is smaller than the standard one, we expect
a significantly smaller multiplicity, i.e., a {\em smaller number
of emitted particles} in the evaporation process, and a {\em
larger average energy} of the produced quanta. Moreover, the BH
stops evaporating when it reaches the minimal mass. This should
lead to the detection of a missing energy of the order of the
minimal mass plus the missing energy due to invisible decay
products.

What happens if $E_{\text{cm}}<M_{\text{min}}$? In this case, the
collision will produce no BH. Either a different nonperturbative
gravitational object (perhaps a brane~\cite{ac} or string
ball~\cite{dec}) forms in the collision, or the scattering is
dominated by nongravitational effects. In the former case,
depending on the details of the quantum gravity theory, we could
have formation of a gravitational object that is either stable or
does not produce any quanta during the decay phase. Therefore, the
formed object could be totally invisible, with the only observable
effect being a missing energy of about the minimal mass in the
final state of the collision.

The minimal BH mass depends sensitively on the (unknown) ${\cal
O}(1)$ parameter $\alpha$ and on the spacetime dimension $d$.
Figure~1 gives the parton-parton $E_{\text{cm}}$ in Planck units
vs.\ the parameter $\alpha$, for $d=6,\dots, 10$ spacetime
dimensions. A BH in $d$ dimensions at fixed $\alpha$ can form only
if $E_{\text{cm}}/M_{\text{Pl}}$ is above the curve corresponding
to the spacetime dimension. For instance, if $\alpha=1$ and
$d=10$, BHs with mass smaller than $\sim 4.72 M_{\text{Pl}}$ do
not form. If $\alpha=2$, the minimum BH mass will be $\sim 600
M_{\text{Pl}}$! In this case, LHC will not see any BH. The
situation is somewhat better in atmospheric ultra high-energy
cosmic ray events, because the collisions are expected to have a
larger $E_{\text{cm}}$ than in particle colliders, and BHs are
expected to form with a higher average mass. However, the bulk of
the BHs which are formed in cosmic ray collisions have mass of the
order of few $M_{\text{Pl}}$, so that the rate of BH formation in
the GUP scenario could be dramatically suppressed. In the worst
case of large parameter $\alpha$, no BHs will be observed, either
at particle colliders or in cosmic ray air-showers.

\begin{figure}\label{figure1}
\begin{center}
\includegraphics[scale=.33, angle=270]{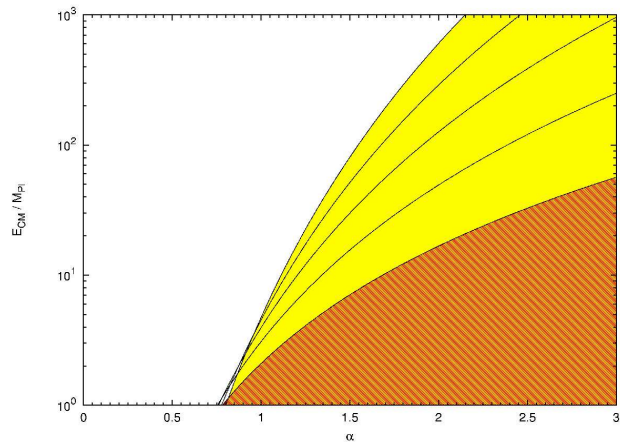}
\caption{Minimal mass for BH formation vs the parameter $\alpha$,
for $d=6,\dots ,10$ dimensions (from below). The yellow and red
regions are the exclusion zones for $d=10$ and $d=6$
respectively.}
\end{center}
\end{figure}

\begin{table*}[t]
\caption{Thermodynamical quantities for two typical
$10$-dimensional BHs produced at LHC (masses $M=9$ and 12~TeV),
assuming $M_{\text{Pl}}=1$ TeV. The values in brackets give the
percentage deviation from standard Hawking quantities.}
\begin{center}
$M=9$ TeV\vskip 12pt
\begin{tabular}{|c||c|c|c|c|c|c|}
\hline &~Min.~mass (TeV)~&~Temperature (TeV)~&~Decay time
(TeV$^{-1}$)~&~Entropy~&~Multiplicity~\\
\hline
$\alpha=0$ &  -- &  .508 &  .689 &  15.5 &  5 \\
$\alpha=0.5$ & .037 & .538 (+6\%)& .511 (-26\%)& 13.3 (-8\%)& 5
(0\%)\\
$\alpha=1.0$ & 4.72 & .721 (+42\%)& .071 (-90\%)& 5.24 (-66\%)& 2
(-60\%)\\
\hline
\end{tabular}\vskip 24pt
$M=12$ TeV\vskip 12pt
\begin{tabular}{|c||c|c|c|c|c|c|}
\hline &~Min.~mass (TeV)~&~Temperature (TeV)~&~Decay time
(TeV$^{-1}$)~&~Entropy~&~Multiplicity~\\
\hline
$\alpha=0$ &  -- &  .488 &  .997 &  21.5 &  7 \\
$\alpha=0.5$ & .037 & .513 (+5\%)& .760 (-24\%)& 20.0 (-7\%)& 7
(0\%)\\
$\alpha=1.0$ & 4.72 & .657 (+35\%)& .157 (-84\%)& 9.62 (-55\%)& 3
(-57\%)\\
\hline
\end{tabular}
\end{center}
\label{numbertable1}
\end{table*}

The region of BH detection at LHC for $d=6$ (yellow) and $d=10$
(red) dimensions is shown in Fig.~2. The upper line correspond to
the CM energy of LHC ($M_{\text{Pl}}=14$ TeV). The two lower lines
correspond to the current experimental limits on $M_{\text{Pl}}$
for $d=6$ ($M_{\text{Pl}}=1.6$~TeV, from submillimeter tests of
the gravitational inverse-square law~\cite{Adelberger:2002ic}) and
$d=10$ ($M_{\text{Pl}}=0.25$~TeV, from collider experiments
\cite{Giudice:2003tu}). The maximum value of the parameter
$\alpha$ that allows observation of BH formation at LHC is
$\alpha\sim 1.6$ ($d=6$) and $\alpha\sim 1.4$ ($d=10$).

\begin{figure}
\begin{center}
\includegraphics[scale=.33, angle=270]{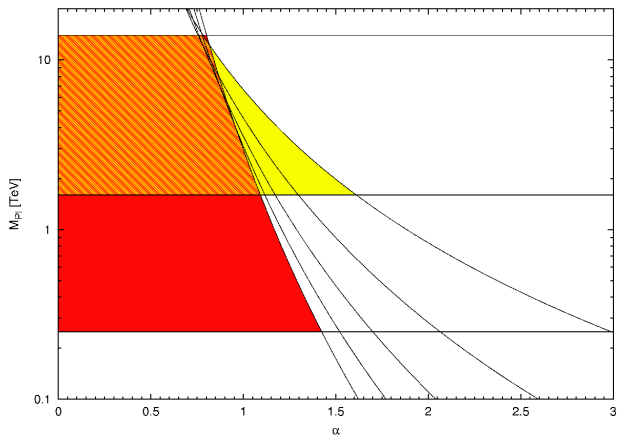}
\caption{($\alpha$,$M_{\text{Pl}}$) moduli space for BH
observation at LHC. The yellow (red) region is for $d=6$ ($d=10$)
spacetime dimensions.}
\end{center}
\label{figure2}
\end{figure}

If the parameter $\alpha$ is sufficiently small, BHs can still be
observed at LHC. If this is the case, what are the observable
signatures of the GUP? We have mentioned that the multiplicity of
the decay process is significantly smaller than in the standard
case. Table~1 shows the parameters for two typical
$10-$dimensional BHs produced at LHC with masses $M=9$ and 12~TeV,
and parameters $\alpha=0$, $0.5$, $1$, assuming
$M_{\text{Pl}}=1$~TeV. The most significant difference between
the standard BH and the GUP-corrected BH is given by the
multiplicity of the decay products. As $\alpha$ increases, the
total number of particles produced during the evaporation phase
becomes smaller. The suppression in the multiplicity is most
severe for $\alpha=1$ and $M=9$ TeV. Indeed, as $\alpha$ increases
and $M$ decreases, the minimum BH mass $M_{\text{min}}$ becomes a
significant fraction of the BH mass $M$. Therefore, a smaller
amount of energy can be converted into Hawking thermal radiation.

The decrease in multiplicity can lead to important observational
signatures. The hadron to lepton ratio of the decay products is
roughly expected to be 5:1. In the case $M=12$~TeV, $\alpha=0$,
for instance, the BH evaporates into five quarks, one charged
lepton, and one gluon, with the charged leptons accounting for
about 15\% of the emission. As the parameter $\alpha$ increases
the leptonic component becomes negligible. When $\alpha=1$, for
example, the charged leptonic component is reduced to $\sim 6$\%
of the emission.

What happens after Hawking evaporation has stopped? The BH remnant
becomes invisible to the detector. This leads to a large missing
energy. (For $\alpha=1$ the missing energy of the 9~TeV BH is more
than 50\% of the total CM energy.) Therefore, an event with
characteristics of a standard BH event (large visible transverse
energy and high sphericity), but with smaller multiplicity and
large missing energy, would be the GUP smoking gun.

In summary, the generalized uncertainty principle of string theory
or non-commutative quantum mechanics brings important qualitative
and quantitative changes to BH thermodynamics. The BH is hotter,
evaporates more rapidly, and has smaller entropy, relative to the
standard case. BH evaporation involves a smaller number of
particles with higher average energy. And the endpoint of
evaporation is a Planck-scale remnant with zero heat capacity. We
have shown, in the context of TeV-scale higher-dimensional
gravity, that these changes could have significant implications
for the possible formation and detection of BHs at LHC (and also
in cosmic ray showers). In particular, the minimum energy for
formation is increased, and could be pushed beyond the reach of
LHC even if $M_{\text{Pl}}\sim 1~$TeV.

~\\ {\bf Acknowledgements}

We thank R.~Bhaduri, P.~Chen, J.~Gegenberg, V.~Husain, M.~Maggiore
and J. Waddington for useful comments and discussions. M.C. and R.M.
are supported by PPARC (UK), and S.D. by NSERC (Canada).

\end{document}